\documentclass{iopconfser}
\usepackage{graphicx}
\usepackage{amsmath}
\usepackage{cite}
\usepackage{hyperref}
\usepackage{fancyhdr}
\usepackage{amssymb}
\providecommand{\keywords}[1]
{
  \textbf{\textit{Keywords:  }} #1
}

\begin{document}
\title{An extensive search for stable periodic orbits of the equal-mass zero angular momentum three-body problem}

\author{Ivan Hristov$^{1}$, Radoslava Hristova$^{1}$,  Kiyotaka Tanikawa$^{2}$}

\affil{$^{1}$ Faculty of Mathematics and Informatics, Sofia University, 5 James Bourchier blvd.,
Sofia 1164, Bulgaria.}

\affil{$^{2}$ National Astronomical Observatory of Japan,
Mitaka, Tokyo 181-8588, Japan.}

\email{ivanh@fmi.uni-sofia.bg,  radoslava@fmi.uni-sofia.bg,  yonoranseonyu@gmail.com} 

\begin{abstract}
 A special 2D initial conditions' domain of the equal-mass zero angular momentum planar three-body problem, which has been formerly studied, is analyzed to deepen the knowledge of the stability regions in it. The decay times in the domain are carefully computed. Four stability regions are established. 971 verified initial conditions for linearly stable periodic collisionless orbits are found. Many of these identified initial conditions are new ones.
 The periodic orbits of each stability region are characterized by a certain pattern in their syzygy sequences. 
 Additional computations show that the orbits found should be considered as candidates for KAM-stable orbits.
\end{abstract}

{\keywords{three-body problem, stable collisionless periodic orbits, high precision computations}}

\section{Introduction}
\label{s:Introduction}

The study of the stability of the classical gravitational three-body problem is of great importance \cite{Marchal:1990, Valtonen:2006},
both in regards to pure mathematics and to astronomy.
The insightful 2009 paper \cite{Martynova:2009} by Martynova, Orlov and Rubinov 
investigates the stability of a special 2D initial conditions' domain of the equal-mass zero angular momentum planar three-body problem.
The main motivation for \cite{Martynova:2009} was that the three well-known stable periodic orbits - the Schubart orbit \cite{Schubart:1956}, Broucke orbit \cite{Broucke:1979} and Figure-8 orbit \cite{Moore:1993, Chenciner:2000, Montgomery:2001} belong to this domain. The first orbit is a rectilinear one, while the second is an isosceles one. These two orbits have collisions and their initial conditions lie at the boundaries of the  domain. 
The Figure-8 orbit is collisionless  and is inside the domain. 

The decay times of the system are computed first in \cite{Martynova:2009} by simulating the motion up to a very long (but limited) time. 
A certain criterion for decay is used. Then the regions of stability are defined by the condition that the solution  is bounded  up to the limited time (the criterion for decay is not satisfied until then). Each of the main three obtained stability regions in \cite{Martynova:2009} include one of the mentioned above three orbits. The stability region generated by the Figure-8 orbit obtained in \cite{Martynova:2009} is qualitatively very similar to the one computed previously in \cite{Simo:2000}.
Although \cite{Martynova:2009} does not use a specialized numerical method in order to capture and compute periodic orbits with good accuracy, the analysis of selected points of the obtained stability regions allows the authors to determine (predict) how the periodic orbits in the  different stability regions generally look. An important result of \cite{Martynova:2009} is the discovery of the simple and stable S-orbit despite the fact that it has been roughly computed. The S-orbit  is a periodic orbit with collisions inside the domain and is classified in \cite{Martynova:2009} as a ``disrupted'' Schubart orbit.

According to the Kolmogorov–Arnold–Moser theory (KAM-theory), 
stable periodic orbits are surrounded by sets of orbits  with nonzero measure, for which the motion is bounded \cite{Arnold:2006}. 
By analyzing the results of \cite{Martynova:2009}, one can assume and predict that the rather large stability regions found are generated by a large number of stable periodic orbits. Indeed, in 2013, a major breakthrough was made in \cite{Suvakov:2013} by ~\v Suvakov and ~Dmitra\v sinovi\'c, when 4 new linearly stable collisionless periodic orbits were discovered in the stability regions from \cite{Martynova:2009}, namely the Bumblebee, Moth I, Moth II and Butterfly III orbits. The work \cite{Suvakov:2013}, together with the subsequent work \cite{Suvakov:2014}, present a detailed and clear description of the numerical search procedure and the topological classification of the orbits. In fact, \cite{Suvakov:2013, Suvakov:2014} broaden the possibility for further searches, including searches of other authors. 
In a later (2018) paper \cite{Veljko:2018}, 16 linearly stable orbits were added to the above-mentioned 4 orbits, thus a list of 20 new orbits was formed. The second known stable choreography after Figure-8 \cite{Suvakov:2014b} is also included in the list.  Meanwhile, in 2017 Li and Liao made an important step in improving the search algorithm through the use of high precision arithmetic \cite{Li:2017}. By conducting a search for relatively longer periods than in \cite{Suvakov:2013, Suvakov:2014},  hundreds of new periodic orbits were discovered in the same domain \cite{Li:2017}. The linear stability of the orbits in \cite{Li:2017} was not investigated in the paper, and  as far as we know such investigation can not be found in any subsequent papers either. Perhaps, this is the reason why the newly-found linearly stable orbits in \cite{Li:2017} (that we conclude to be at least 101 in our current paper) were not included in the subsequent work \cite{Veljko:2018}. Later prompted by the search in \cite{Suvakov:2014b}, more than 100 new linearly stable topological powers of Figure-8 with relatively long periods were discovered \cite{Hristov:2023, Hristov:2022}. Overall, although the periodic orbits corresponding to the stability region for the Figure-8 orbit are a relatively well-examined topic, it is not so for the other regions.

We have to mention the thorough work of Danya Rose \cite{Rose:2016}. This work successfully implements the numerical methods developed in \cite{Rose:2013}. Similar to our research, equal-mass zero angular momentum planar periodic orbits are studied in it, but some periodic orbits outside the 2D domain considered by us are also found. Since it was not easy for us to separate these orbits, we did not  manage to strictly check whether we have missed some orbits from \cite{Rose:2016}. Anyway, the linearly stable collisionless periodic  orbits in \cite{Rose:2016} are 25 in number (see Table 6.6 in \cite{Rose:2016}) and  the considered scale-invariant periods are shorter than ours. A part of these 25 orbits are discovered independently and included in \cite{Veljko:2018}. \cite{Rose:2016} is particularly valuable due to its discovery of stable periodic orbits with collisions. The S-orbit (t2(8,2)), as well the Broucke orbit (isosceles (8,1)) and the Schubart orbit (rectilinear (16,1)) are rediscovered and analyzed. Their linear stability is shown through precise computations. Let us also note that Broucke is the only isosceles stable orbit found, and Schubart is the only rectilinear stable orbit found in \cite{Rose:2016}.

With this paper we intend to confirm and expand the knowledge about this interesting 
2D initial conditions' domain through our high precision and verified numerical results.
We set ourselves the following main goals :

$\bullet$ To present high precision results for the decay times in the domain.

$\bullet$ To try to gather all known linearly stable collisionless periodic orbits from the considered
domain in one place and to conduct an extensive long periods' search for such orbits in order to expand the list of known ones.

$\bullet$ To present the initial conditions and the periods of the found linearly stable orbits with many correct digits .

$\bullet$ To perform a high precision  study of the linear stability and provide the eigenvalues of the monodromy matrices
with many correct digits.

$\bullet$ To check if the non-resonance condition from the KAM-theory is satisfied for all found orbits.

$\bullet$ To establish  a correspondence between the different stability regions found and a certain pattern 
in the syzygies sequences of the stable periodic orbits in these regions.

The paper is organized into 6 sections.
After the introduction, in Sect. \ref{s:Mathematical model} we present the differential equations governing the motion of three bodies and the initial conditions. In Sect. \ref{s:Numerical search details} we give details about the conduct of the search and the numerical methods used.
In Sect. \ref{s:Studying stability} we explain how the stability study is done. The results are presented in Sect. \ref{s:Results}. We draw the conclusions and define the outlook in Sect. \ref{s:Conclusions}.

\section{Mathematical model}
\label{s:Mathematical model}
\subsection{Differential equations}
\label{ss:Differential equations}
The ordinary differential equations governing the motion of three bodies are derived
from Newton's second law and Newton's law of gravity:
\begin{equation}
\label{equasecond}
m_i\ddot{r}_i=\sum_{j=1,j\neq i}^{3}G m_i m_j \frac{(r_j-r_i)}{{\|r_j -r_i\|}^3}, i=1,2,3.
\end{equation}
In (\ref{equasecond}) $G$ is the gravitational constant, $m_i$ are the masses and $r_i(t)$ are the vectors of the positions in space. 
The derivatives are with respect to the time $t$.
The model treats the bodies as point masses. 
We consider $G=1$ and the masses to be equal: $m_1=m_2=m_3=1$.
In this work we also consider a planar zero angular momentum  three-body motion.
So the vectors $r_i(t)$, $\dot{r}_i(t)$ have two components: $r_i=(x_i, y_i)$,
$\dot{r}_i=(\dot{x}_i, \dot{y}_i).$
The dependent variables ${vx}_i$ and ${vy}_i$ are introduced, so that ${vx}_i=\dot{x}_i, {vy}_i=\dot{y}_i.$
Then, the second order system (\ref{equasecond}) can be written as a first order one:
\begin{equation}
\label{equations}
\dot{x}_i={vx}_i, \hspace{0.1 cm} \dot{y}_i={vy}_i,  \hspace{0.1 cm} \dot{vx}_i=\sum_{j=1,j\neq i}^{3}\frac{(x_j-x_i)}{{\|r_j -r_i\|}^3},  \hspace{0.1 cm} \dot{vy}_i=\sum_{j=1,j\neq i}^{3}\frac{(y_j-y_i)}{{\|r_j -r_i\|}^3}, \hspace{0.1 cm} i=1,2,3.
\end{equation}
We use this first order form of the system, because it is more suitable for analytical and numerical studies. 

\begin{figure}
\centerline{\includegraphics[width=0.5\columnwidth,,keepaspectratio]{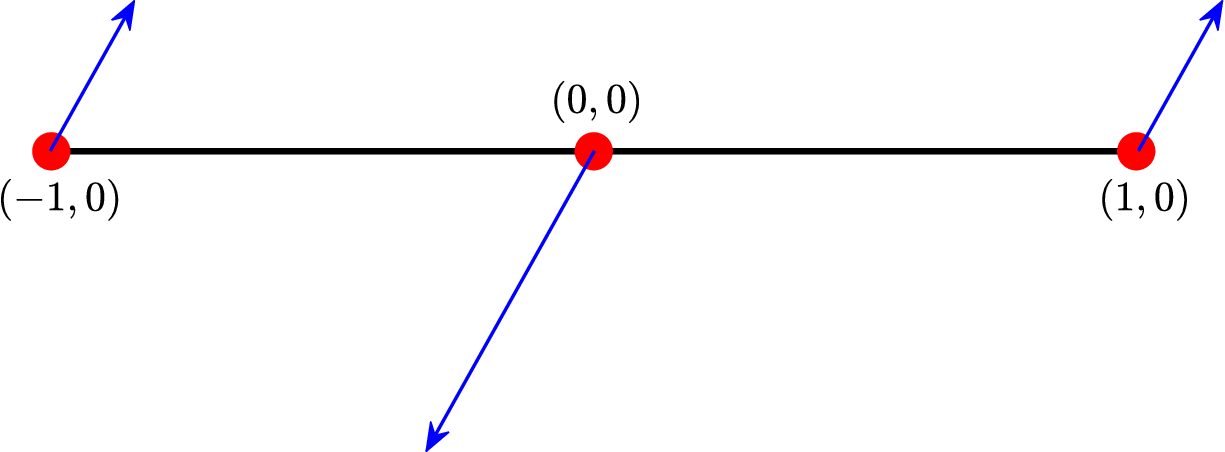}}
\caption{\footnotesize{Initial conditions}}
\label{fig:Euler}
\end{figure}

\subsection{Initial conditions}
\label{ss:Initial conditions}
The initial conditions are determined by two parameters $v_x\geq0$, $v_y\geq0$ as in \cite{Suvakov:2013, Suvakov:2014, Li:2017, Hristov:2024}: 
\begin{center}
\begin{equation}
\label{initcond}
\begin{split}
(x_1(0),y_1(0))=(-1,0), \hspace{0.2 cm} (x_2(0),y_2(0))=(0,0), \hspace{0.2 cm} (x_3(0),y_3(0))=(1,0) \hspace{0.5 cm} \\
({vx}_1(0),{vy}_1(0))=({vx}_3(0),{vy}_3(0))=(v_x,v_y)  \hspace{2.5 cm}\\
({vx}_2(0),{vy}_2(0))=-2({vx}_1(0),{vy}_1(0))=(-2v_x, -2v_y) \hspace{1.8 cm}
\end{split}
\end{equation}
\end{center}

\begin{figure}
\centerline{\includegraphics[width=0.5\columnwidth,,keepaspectratio]{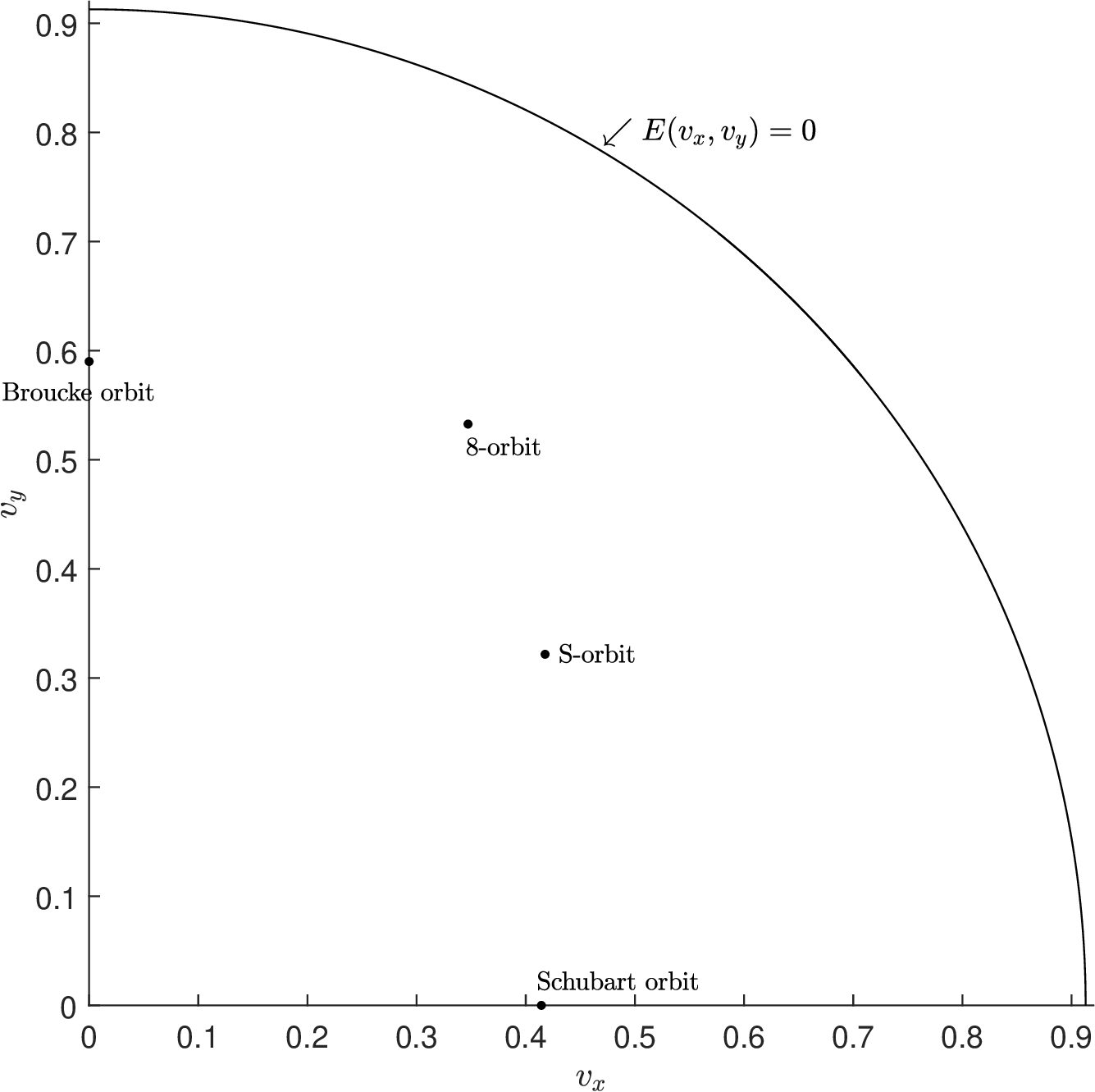}}
\caption{The 2D initial conditions' domain bounded by $v_x=0$ and $v_y=0$ axis and $E=0$ curve.}
\label{fig:search domain}
\end{figure}

We call these initial conditions Euler half-twist initial conditions, as called in \cite{Hristov:2024}. One can see a picture of the initial positions and velocities (with arrows) in Figure \ref{fig:Euler}. The initial conditions have the following symmetries. The three bodies lie on a line (are in syzygy) at $t=0$ and the second body is in the middle between the other two (is at the center of mass). The velocities of the first and third bodies are the same, while the velocity of the second body is twice the velocities of the others and has an opposite direction. The known  Schubart, S, Broucke and Figure-8 orbits  share the property of passing at some instance through such initial conditions. For  the Schubart orbit $v_y=0$ and for the Broucke orbit $v_x=0$. These initial conditions have  zero linear and zero angular momentum and have energy $E(v_x, v_y)=-2.5+3({v_x}^2+ {v_y}^2).$ As negative energy is necessary for a bounded motion \cite{Suvakov:2014},
the 2D search domain is those bounded  by the $v_x=0$ and $v_y=0$ axis and the $E(v_x, v_y)=0$ curve (see Figure \ref{fig:search domain}).
The Schubart and Broucke orbits are on the boundaries.
It is also important to note that the two parameters for the same 2D initial conditions' domain are chosen differently in \cite{Martynova:2009}. We prefer the choice in \cite{Suvakov:2013, Suvakov:2014, Li:2017, Hristov:2024}, because it is most convenient for our numerical computations.

From the above symmetries it follows that if a periodic orbit has Euler half-twist initial conditions at $t=0$, then it has Euler half-twist initial conditions once again at $t=T/2$, where $T$ is the period. One can see the proof of this property given by R. Montgomery in the appendix of \cite{Hristov:2024}. We use this symmetry property in order to conduct a more efficient numerical search (see \cite{Hristov:2024} for details). 

\section{Numerical search details}
\label{s:Numerical search details}

To search for periodic orbits we use Newton's method with initial approximations obtained by the grid-search method \cite{Abad:2011}. 
To overcome the problems stemming from the sensitivity on the initial conditions and to compute reliable long term solutions,
we make use of high precision computations.
We apply the high order Taylor Series Method (TSM) used with high precision arithmetic as an ODE-solver
\cite{Jorba:2005, Barrio:2006, Barrio:2011, Izzo:2021}. The Taylor coefficients are computed using the rules of automatic differentiation \cite{Jorba:2005, Barrio:2006, Barrio:2011, Izzo:2021}. The 2017 work \cite{Li:2017} demonstrates the great success of Newton's method combined with a high precision TSM for the three-body problem. In our recent work \cite{Hristov:2024}, we improved the efficiency of the search algorithm from \cite{Li:2017}  by solving the equation expressing the property that at the half period the periodic solutions have Euler half-twist
initial conditions again. Here, we use this more efficient approach. Solving a half-period equation instead of the standard full period equation, is critical for the search for periodic orbits in the chaotic regions \cite{Hristov:2025}. This effect is explained in \cite{Hristov:2025} in the terminology of the Lyapunov exponents of the unstable periodic orbits. Furthermore, this also benefits the stability regions, at least because the integration time is divided by two, which at least allows the use of a finer search-grid for the same computational work. Note, however, that our methods are not designed for periodic orbits with collisions, although they can capture and roughly approximate such orbits.

Details about Newton's method applied to the initial conditions' domain under consideration, including the search stages, the choice of the time step-size, the choice of Taylor's order and the precision, can be found in \cite{Hristov:2024}. The formulas for automatic differentiation can be found in the previous work \cite{Hristov:2021}. We just want to note that for the time step strategy we apply, we make the optimal choice to use  64 bit of precision for every 22 Taylor's order. 

The opportunity to choose arbitrary order and precision of TSM allows the application of a verification procedure based on TSM which produces reliable solutions and is called Clean Numerical Simulation (CNS) \cite{Li:2018}. The potential of CNS goes far beyond the three-body problem, as it can also be extended to PDEs \cite{Hu:2020, Qin:2020, Liao:2023}.

In order to search for stable periodic orbits, we first introduce a quadratic grid
with a size of  $2^{-9}$ ($\approx 171000$ points) inside the domain in
Figure \ref{fig:search domain}. Then, we compute the decay time at each point by integrating the system until the condition for decay is satisfied or until the scale-invariant time $t^*=t|E|^{3/2}$ reaches 10000. The decay condition is as in \cite{Martynova:2009}, i.e. the maximal distance between the bodies $r_{max}>5d$, where $d=3/|E|$ is the average triple system size. After that, around the points for which the scale-invariant decay time $t^*_{decay} >1500$, we introduce a grid of points with a size of $2^{-13}$ ($\approx 7000000$ points). Next, we conduct a search at these points for periodic orbits with scale-invariant periods $T^*=T|E|^{3/2} < 800$. These are relatively long periods. About 95000 periodic orbits are found at this stage. 
We compute these orbits' initial conditions and periods with an accuracy corresponding to 128 bit of precision.
Afterwards, we check whether the solutions diverge (depart) from the periodic orbit after a 100 periods' simulation with 128 bit of precision and leave those solutions that do not diverge. They become our candidates for stable periodic orbits. Following this, we increase the precision and the order of TSM so that Newton's method converges up to a return proximity $10^{-220}$ (see the notion of return proximity in \cite{Hristov:2024}). We also check if when we start with 100 correct digits approximations of the  initial conditions and periods  the ratios of the logarithms of the return proximities for two successive Newton's iterations are approximately 2 - the theoretical order of convergence of Newton's method.  This check of the regular convergence of Newton's method up to such small return proximities and computing the orbits with hundreds of correct digits gives us much greater confidence in the existence of the periodic orbits found, and it also gives us highly accurate data with which to start our stability study. 

\section{Studying stability}
\label{s:Studying stability}

In this work we do a linear stability study, firstly,  by computing the monodromy matrices
for the corresponding periodic solutions and then computing their eigenvalues (see \cite{Veljko:2018, Hristov:2024b} for details).
The eigenvalues are scale-invariant quantities.
Note that although the term ``linear stability'' is widely used for the three-body problem and we use it generously here, 
it is more correct to use the term ``spectral stability''. See Sect. 2.4 in \cite{Montgomery:2025} for the different types of stability - their definitions and the relations between them. 
The computation of the monodromy matrices does not create additional technical difficulties because they are of the same Jacobian type as the matrices in Newton's method and are computed in the same way - through TSM. They are computed with 160 correct digits.
Four of the eigenvalues determine the linear stability. The remaining eigenvalues are all equal to one \cite{Roberts:2007}.
We are interested in the case, in which we have elliptical stability, i.e. when the four eigenvalues that determine the stability 
are two complex conjugate pairs on the unit circle:
$$ e^{i 2 \pi \omega_1}, e^{-i 2 \pi \omega_1}, e^{i 2 \pi \omega_2}, e^{-i 2 \pi \omega_2}, \quad  0<\omega_1<\omega_2<0.5.$$
In this case we call the solution linearly stable. The real numbers $\omega_1$ and $\omega_2$ are called frequencies \cite{Montgomery:2025} or sometimes stability angles or
Floquet exponents \cite{Veljko:2018}.
In this work we present verified results for the eigenvalues and the frequencies $\omega_1, \omega_2$ by comparing the results of two computations with different precisions. 
The eigenvalues are computed with the Multiprecision Computing Toolbox \cite{Advanpix} for MATLAB\textregistered \cite{Matlab}.
We want to mention that all given digits of  $\omega_1$ and $\omega_2$ for the Figure-8 orbit from \cite{Simo:2000} coincide with our first digits.
To be confident in our results, we discard those orbits from the list, for which there is a pair of eigenvalues that are very close to 1 or -1 and that do not pass our verification test. Even our high-precision calculations cannot confirm with certainty whether these solutions are linearly stable or not. Some of these discarded solutions will turn out to be elliptically stable, but others will be hyperbolic-elliptic (one hyperbolic and one elliptic eigenvalues pair). 
Note that some of the hyperbolic-elliptic orbits with a small Lyapunov exponent pass the initial rough test of a simulation of 100 periods. After we separate all linearly stable orbits, we do a test with only 64 bit precision and simulation up to $t^*=10^6$ and none of the solutions found diverge from the periodic orbit during this rather long time.

When we use the shortened term ``stability'' in this paper, we most often mean ``linear stability''.
The ultimate study of the stability of the found periodic orbits, however, should be for KAM stability.
To meet the KAM stability, not only must the necessary linear stability be fulfilled, but also the non-resonance condition and the twist condition must be added (see Sect. 2.4 in \cite{Montgomery:2025} and the corresponding theorem). 
The non-resonance condition states that we have no resonances of order 4 or less, i.e. there is no nonzero integer vector solution  $(k_0, k_1, k_2)$ to
the linear equation $k_1\omega_1+k_2\omega_2=k_0$ having $0<|k_1|+|k_2|\leq 4$ (see again Sect. 2.4 in \cite{Montgomery:2025}).
The check of the non-resonance condition is not difficult, if we have a verified result for $\omega_1$ and $\omega_2$ and this check is done here.
Checking of the twist condition remains for future work.

\section{Results}
\label{s:Results}

\subsection{A colour map of the decay times}
\label{ss:Colour map of decay times}

\begin{figure}[h!]
\centerline{\includegraphics[width=1.0\columnwidth,,keepaspectratio]{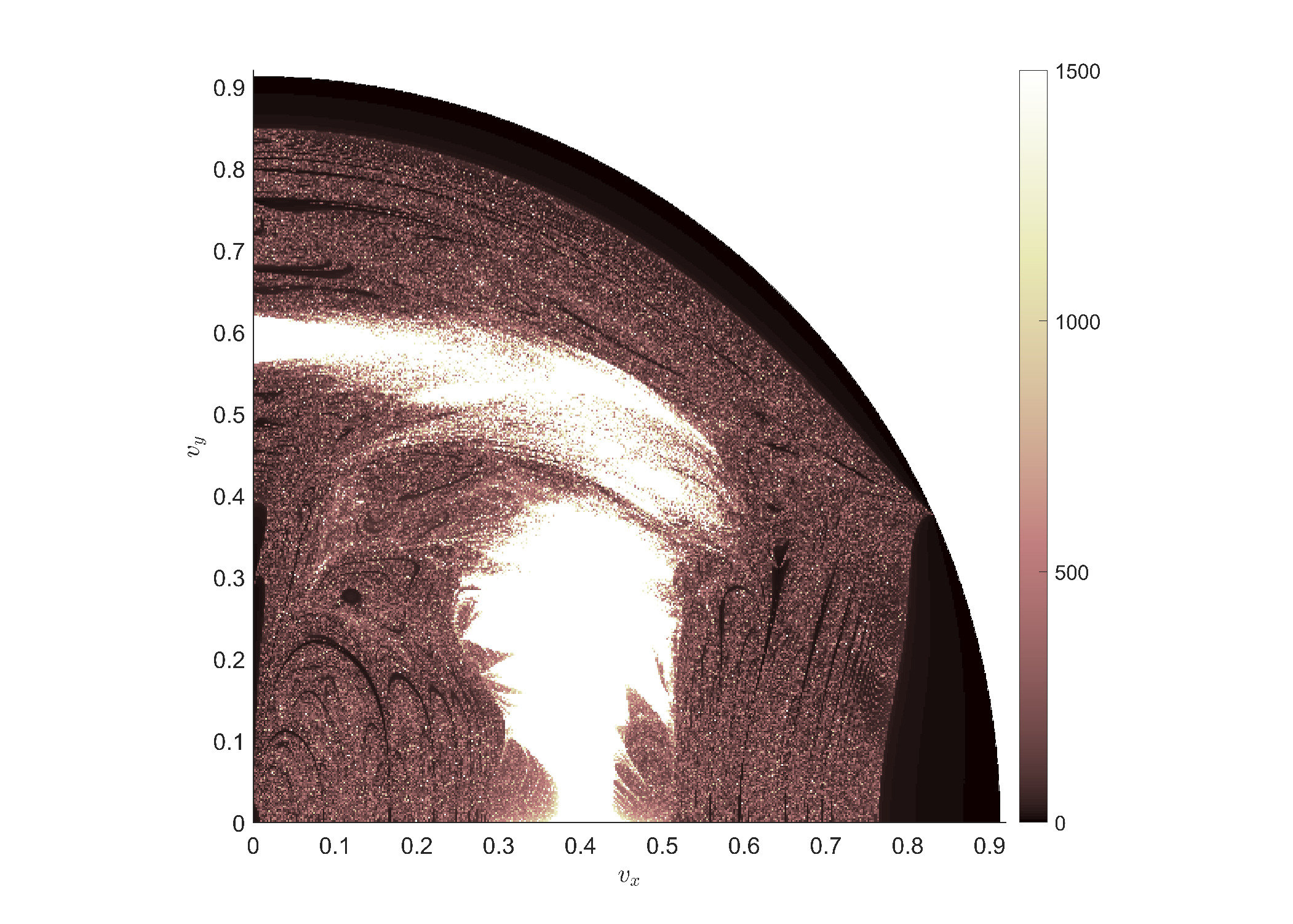}}
\caption{A colour map of the decay times.}
\label{fig:decay time}
\end{figure}

 The decay times in the 2D initial conditions' domain are used to determine the regions, in which we are to look for stable periodic orbits. This study, however, is also of its own interest. Figure \ref{fig:decay time} presents a colour map of the decay times.
 We compute the scale-invariant decay times with a simulation up to $t^*=10000$, but the colour bar in Figure \ref{fig:decay time} is up to 1500, because we judged that this way we get a more expressive picture of the colour map, and also, it is precisely around the points with $t^*_{decay}>1500$ that we are looking for stable periodic orbits. In dark brown, in Figure \ref{fig:decay time} are the regions with very short decay time, in which there are no periodic orbits. In white, are the stability regions (the points with $t^*_{decay}>1500$). The remaining regions, in inhomogeneous light brown, are the chaotic regions. Our results for the decay times are qualitatively very similar to those in \cite{Martynova:2009}.
Note that since we integrate only up to finite times, the regions of stability only give us an idea of where the true regions of bounded motion are contained. We will give in the next subsection the smaller stability regions for $t^*_{decay}>10000$, which is a better approximation of the real bounded motion' regions.

We verify our decay times' results with our high precision ODE-solver TSM. We do this by comparing the results for gradually increased precision  and also with a reversibility test. As the precision increases, the unverified points decrease, until all of them are finally verified. 
In fact, we apply a form of CNS. This is a rather expensive numerical procedure, due to the need for very high precision. Thus we cannot process too many points in the domain.  The points for the colour map presented here are $\sim 171000$.

We want also to mention that most of the periodic orbits found in \cite{Suvakov:2013, Li:2017} are in the stability regions, despite the fact that
the entire domain is considered. One may be left from these papers with the impression that the periodic orbits are concentrated precisely there. Actually, the percent of the stable orbits in \cite{Suvakov:2013, Li:2017} is rather high, because the numerical approaches used are not well-suited for unstable orbits. This can of course have its advantages, because with a small risk of missing the important stable periodic orbits, computational resources are not wasted in order to process unstable orbits. In  \cite{Hristov:2024}, however, it is shown that when using the half-period approach, many (hundreds of thousands) unstable periodic orbits  can be found in the chaotic regions. 

\subsection{Distribution of the initial conditions (i.c.s) found and their syzygy sequences}
\label{ss:initial conditions found}

The main result of the current work is the finding of 971 linearly stable collisionsless periodic orbits' i.c.s with $T^*<800$. 
From this result follows that the stability regions in the initial conditions' domain  seem to be generated by 
many (possibly infinitely many) stable periodic orbits. 
All orbits studied for linear stability from the previous works \cite{Veljko:2018, Hristov:2023, Hristov:2022} 
are rediscovered and included in the new list. 
We checked that the list of orbits from \cite{Li:2017} contains at least 101 new linearly stable orbits, which were also not missed. 
We say ``at least'', because it is possible that there are other linearly stable orbits that have not passed our verification tests.
The rediscovery of the previously known by us stable periodic orbits confirms the good quality of the search conducted.

\begin{figure}[h!]
\centerline{\includegraphics[width=0.8\columnwidth,,keepaspectratio]{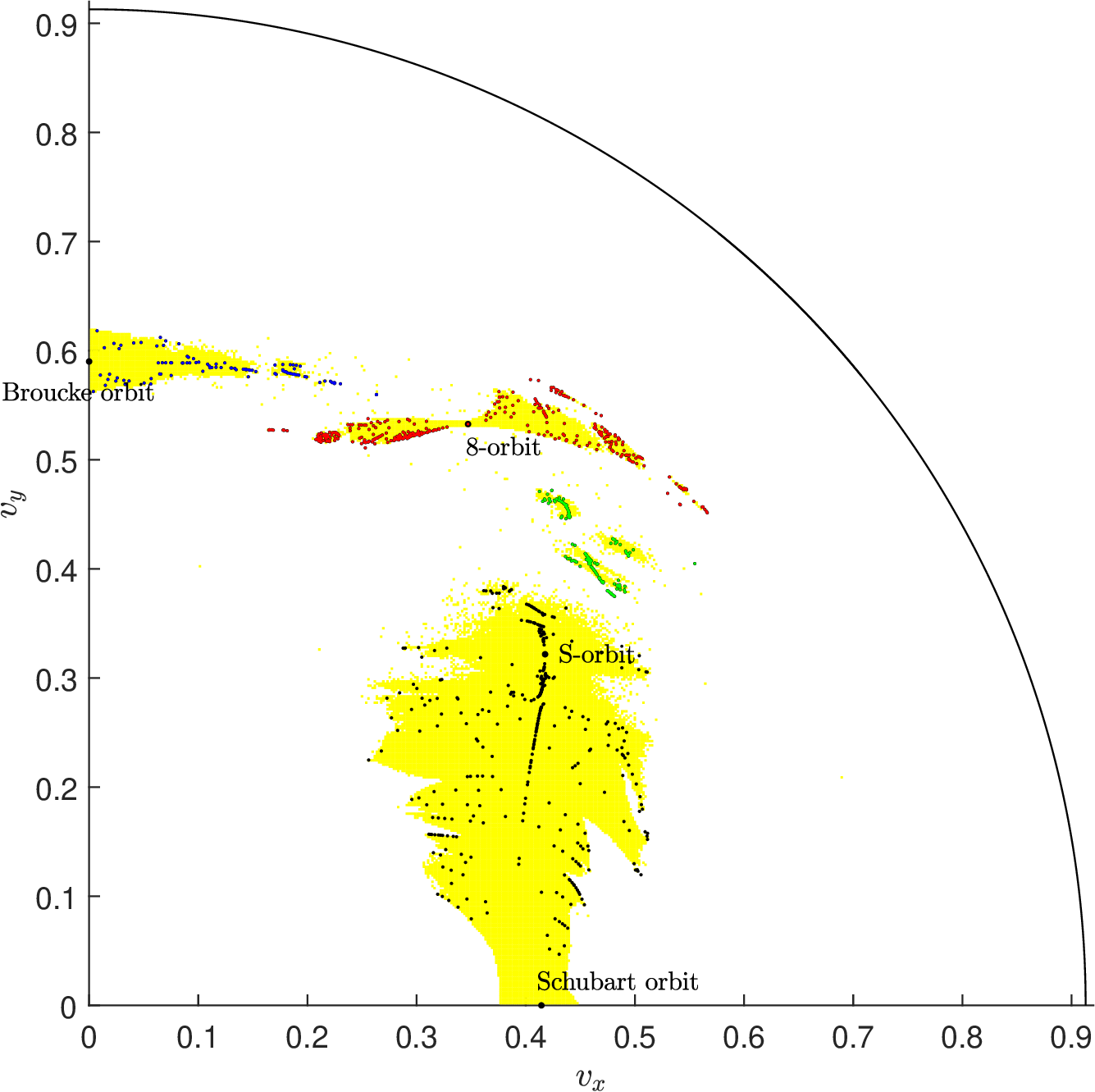}}
\caption{Distribution of the stable i.c.s. The stability regions ($t^*_{decay}>10000$) are in yellow.}
\label{fig:distribution}
\end{figure}

Figure \ref{fig:distribution} shows the distribution of the found i.c.s. As in \cite{Martynova:2009} we have four stability regions (in yellow).
The yellow regions correspond to the points from the grid with a size $2^{-9}$ for which $t^*_{decay}>10000$. It can be assumed that these yellow regions are a good approximation of the real regions with bounded motion. Surely, some very small yellow regions are not detected, due to the insufficiently fine grid, but remember that we search for periodic orbits around the points with a relaxed stability condition - points with $t^*_{decay}>1500$.  The largest region is generated by the Schubart and  S orbits (377 i.c.s). We have one region generated by the Figure-8 orbit (346 i.c.s) and one region generated by the Broucke orbit (101 i.c.s). We also have a small intermediate region (of small islands) between the Figure-8 region and the largest region with 147 i.c.s. Observations on the trajectories of the stable periodic orbits found show that they generally reproduce the main features of the four ``basic'' orbits (Schubart, S, Figure-8, and Broucke orbits). 
It seems that the Figure-8 orbit plays a constructive role in each other region by ``hybridizing'' itself with the
``basic'' orbit from the corresponding region, so that in many cases each body moves along a trajectory consisting of ``distorted'' eights.
Let us note again that although the work \cite{Martynova:2009} does not use a tool for accurate computing of periodic orbits, the general form of the periodic orbits' trajectories in the different stability regions is well predicted (see Figures 7, 8, 9, 10 and their description in \cite{Martynova:2009}). 

\begin{figure}[h!]
  \centerline{\includegraphics[scale=0.52]{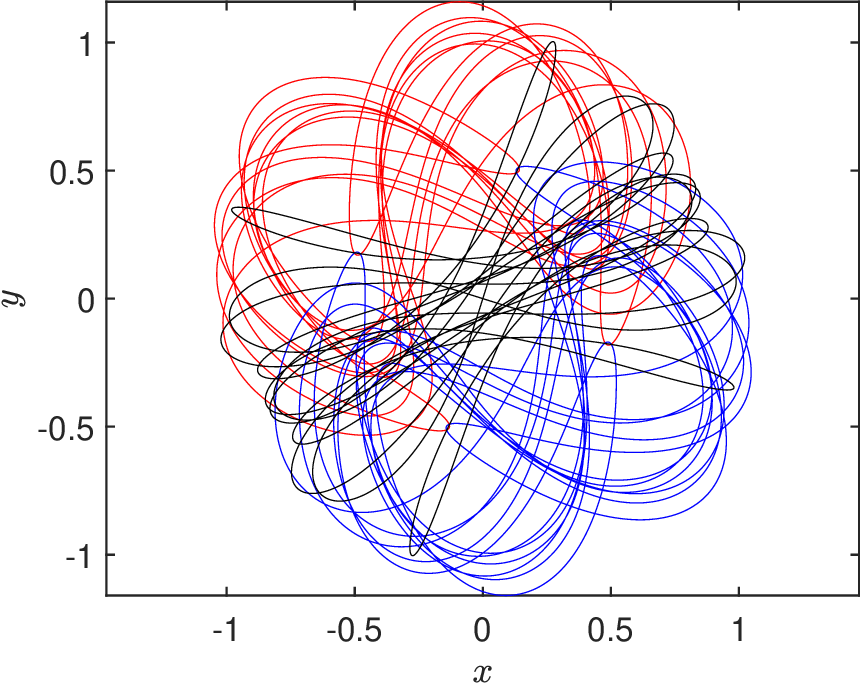}
  \hspace{0.5 cm}
  \includegraphics[scale=0.52]{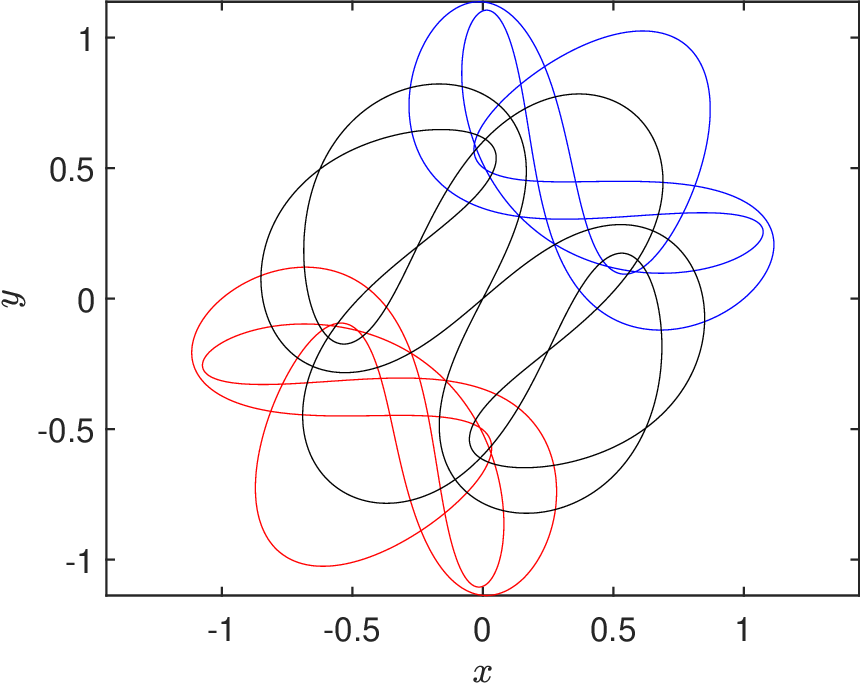}}
  \caption{\small{Bumblebee \cite{Suvakov:2013} -- left and Moth I \cite{Suvakov:2013} -- right}}
  \label{trajectories1}
\end{figure}

\begin{figure}[h!]
  \centerline{\includegraphics[scale=0.52]{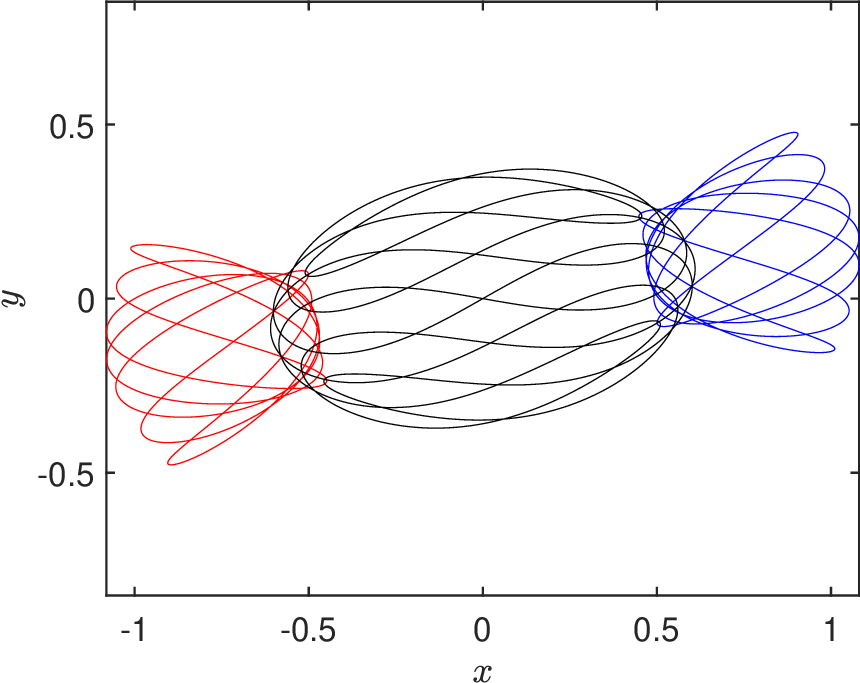}
  \hspace{0.5 cm}
  \includegraphics[scale=0.52]{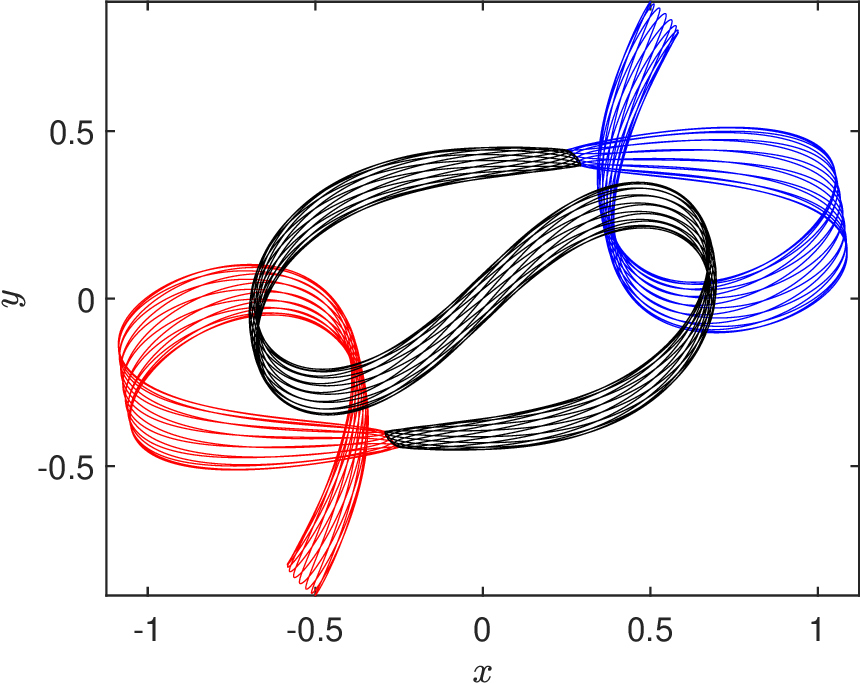}}
  \caption{\small{Butterfly III \cite{Suvakov:2013} -- left and a S-like  orbit -- right}}
  \label{trajectories2}
\end{figure}

For each orbit we calculated the syzygy sequences \cite{Montgomery:2019, Tanikawa:2008, Tanikawa:2015} using the numbering of the bodies from the initial conditions (\ref{initcond}). The syzygy sequences in the region of Figure-8 are completely clear - they are of the form $(213)^{2k}, k  \in \mathbb{N}$. 
For the other regions we did not manage to find a general formula, but we searched for some pattern that characterizes the corresponding sequences. We obtained the following results. The orbits in the Broucke region are characterized by containing the subsequence $(213)^2(231)^2$ in their syzygy sequences. These are the blue dots in Figure \ref{fig:distribution}. The most well-known collisionless periodic orbit of this type is Bumblebee \cite{Suvakov:2013}. See its trajectory in Figure \ref{trajectories1} (left). The orbits in the intermediate region with the islands (the green dots) are characterized by the fact that their syzygy sequences contain the subsequence $(213)^2 2123 (123)^2$, considered cyclically, and do not contain $(213\hspace{0.2cm}2123\hspace{0.2cm}123)^2$. The most well-known collisionless periodic orbit here is Moth I \cite{Suvakov:2013} (see Figure \ref{trajectories1} (right)). Finally, all orbits of the largest region (the black dots) are characterized by the condition that their syzygy sequences contain $(213\hspace{0.2cm}2123\hspace{0.2cm}123)^2$ or $(2123\hspace{0.2cm}123)^2$, considered cyclically. The most well-known collisionless periodic orbit here is Butterfly III \cite{Suvakov:2013}, given  in Figure \ref{trajectories2} (left). Figure \ref{trajectories2} (right)  also shows a S-like orbit from the vicinity of the S-orbit. 

\subsection{The available high precision data}
\label{ss:Available data}

The data of the 971 i.c.s (685 distinct solutions) found is presented in the form $(v_x, v_y, T, T^*)$ with 100 correct digits.
It can be found as a text file with 971 rows and 4 columns in \cite{Paper web-site:2025}. The rows are sorted by $T^*$ (the forth column). 
Some solutions (286) are presented by two different i.c.s in the domain. They are given as two adjacent rows with the same fourth column $T^*$. These are solutions of symmetry type II (see \cite{Hristov:2024}). The rest of the solutions (399) are represented by 
only one initial condition (solutions of symmetry type I). The four eigenvalues determining the linear stability and the two stability angles (frequencies) $\omega_1$ and
$\omega_2$ calculated through them are given with 40 correct digits in \cite{Paper web-site:2025}. The syzygy sequences and their lengths are uploaded too.
Real space plots of the trajectories for all i.c.s are drawn. In addition, all i.c.s, syzygy sequences and plots are divided into four groups according to their stability region and one can consider each stability region separately.  
There are also several animations.

The computation of the stability angles $\omega_1$ and $\omega_2$  
with 40 correct digits allows us to check the non-resonance condition for KAM-stability easily.
All i.c.s satisfy it. This along with the very long simulation up to $t^*=10^6$, for which the solutions remain close to the orbit obtained after the first loop, turns all found i.c.s into candidates for KAM-stable ones. What remains is to verify the twist condition as it is done for the Figure-8 orbit in \cite{Simo:2000}.
It would be excellent to prove the KAM-stability rigorously as done for the Figure-8 in \cite{Kapela:2007}, at least for some i.c.s. We hope that our high precision results will motivate such studies being carried out.

The extensive computations are made in the Nestum cluster, Sofia Tech Park, Sofia, Bulgaria \cite{Nestum}.

\section{Concluding remarks and outlook}
\label{s:Conclusions}
$_{}$

$\bullet$ A special formerly studied 2D initial conditions' domain for the equal-mass zero angular momentum planar three-body problem is considered.
971 verified linearly stable  collisionsless periodic orbits' i.c.s are found and given with 100 correct digits.
A large part of the found i.c.s are new ones.

$\bullet$ Verified decay times' results are presented as a colour map.

$\bullet$  All found linearly stable i.c.s satisfy the non-resonance condition for KAM-stability.
The complete KAM-stability study is a matter for future work.

$\bullet$ There are four stability regions established in the initial conditions' domain. 
It is shown that the solutions from each region are characterized by a certain pattern in their syzygy sequences.
The syzygy sequences should be studied in greater detail with the ultimate goal of deriving general formulas for them.

$\bullet$ The generalization with respect to the masses of the considered 2D domain from \cite{Li:2018b}
should continue to be investigated in the future. There is, at the very least, a lack of studies on the decay times for different mass ratios.
The potential of this future study is revealed in the work \cite{Veljko:2023}, the results of which suggest the possible existence of 
exotic circumbinary exoplanets.

$\bullet$ It is worth investigating the established equal-mass stability regions in a meaningful  higher dimensional domain that contains the given domain.

\section*{Acknowledgement}
The authors acknowledge the access to the Nestum cluster, HPC Laboratory, Research and Development and Innovation Consortium, Sofia Tech Park, funded by the National Roadmap for Research Infrastructures: 2018 - 2027.

\end{document}